\documentclass{rmaa}
\usepackage{amsmath}

\title{Velocity segregation in a clump-like outflow with a
non-top hat velocity cross-section}
\author{
  A. Castellanos-Ram\'\i rez$^1$, A. C. Raga$^2$, J. Cant\'o$^1$,
  A. Rodr\'\i guez-Gonz\'alez$^2$, L. Hern\'andez-Mart\'\i nez$^3$
  \affil{$^1$Instituto de Astronom\'\i a, UNAM,
    $^2$Instituto de Ciencias Nucleares, UNAM,
    $^3$Facultad de Ciencias, UNAM}}

\fulladdresses{
\item A. Castellanos-Ram\'\i rez, J. Cant\'o: Universidad Nacional Aut\'onoma de M\'exico, Instituto Astronom\'\i a, Ap. 70-264, CDMX, 04510, M\'exico
\item A. C. Raga, A. Rodr\'\i guez-Gonz\'alez: Instituto de Ciencias Nucleares,
Universidad Nacional Aut\'onoma de M\'exico, Ap. 70-543, 04510 CDMX,
M\'exico (raga@nucleares.unam.mx)
\item L. Hern\'andez-Mart\'\i nez, Facultad de Ciencias, Universidad
  Nacional Aut\'onoma de M\'exico, Av. Universidad 3000, Circuito
Exerior S/N, 04510 CDMX, M\'exico}

\shortauthor{Castellanos-Ram\'\i rez et al.}
\shorttitle{Velocity segregation}

\SetVolume{40} \SetFirstPage{001} \SetYear{2020}
\ReceivedDate{\today} 
\AcceptedDate{Year Month Day}

\resumen{Nudos de alta velocidad unidos a la fuente del flujo
  por una emisi\'on con una rampa de ``ley de Hubble'' de velocidades
  radiales con crecimiento lineal en funci\'on de la distancia son observados
  en algunas nebulosas planetarias y en algunos flujos en regiones de
  formaci\'on estelar. Proponemos un modelo simple en el que un
  ``nudo'' es eyectado de una fuente en un per\'\i odo $\tau_0$, con
  una fuerte estratificaci\'on de velocidades del eje al borde. Esta
  estratificaci\'on resulta en una superficie de trabajo curvada
  (inicialmente empujada por el material eyectado, y luego movi\'endose
  bajo su propia inercia). Tanto en los modelos anal\'\i ticos como
  las simulaciones num\'ericas encontramos que esta superficie
  de trabajo tiene una dependencia lineal de velocidad vs. posici\'on,
  por lo tanto se reproducen cualitativamente los ``nudos
  con ley de Hubble'' en nebulosas planetarias y en flujos de
  estrellas j\'ovenes.}

\abstract{High velocity clumps joined to the outflow source
  by emission with a ``Hubble law'' ramp of linearly increasing radial
  velocity vs. distance are observed in some planetary nebulae
  and in some outflows in star formation regions. We propose
  a simple model in which a ``clump'' is ejected from a source
  over a period $\tau_0$, with a strong axis to edge velocity
  stratification. This non-top hat cross section results in the
  production of a highly curved working surface (initially being
  pushed by the ejected material, and later coasting along due to
  its inertia). From both analytic models and numerical simulations
  we find that this working surface has a linear velocity vs. position
  ramp, and therefore reproduces in a qualitative way the
  ``Hubble law clumps'' in planetary nebulae and outflows
  from young stars.}

\keywords{stars: winds, outflows --
  ISM: jet and outflows -- ISM: Herbig-Haro objects --
  ISM: planetary nebulae}

\begin{document}
\maketitle

\section{Introduction}

In different stellar outflows, one sometimes finds clump-like
flows with an emitting ``trail'' (linking the clumps to the
outflow source) with a ``Hubble law'' of linearly increasing velocities
with distance from the source. This kind of structure is observed
in some outflows from young stars (most notably in the Orion BN-KL
outflow, see e.g. \citet{Allen1983}, \citet{Bally2017} and
\citet{Zapata2011} and in some planetary (PN) and
protoplanetary (PPN) nebulae (see, e.g., \citealt{Alcolea2001}
and \citealt{Dennis2008}). A second, striking outflow with
multiple ``Hubble tail clumps'' has been recently found
by \citet{Zapata2020}.

Following the suggestion of \citet{Alcolea2001} that
the observed ``Hubble law tail'' clumps were the result
of ``velocity sorting'' of a sudden ejection with a range
of outflow velocities, \citet{Raga2020a,Raga2020b} developed
a model of a ``plasmon'' resulting from a ``single pulse''
ejection velocity variability. In this model, an ejection
velocity pulse of parabolic \citep{Raga2020a} or
Gaussian \citep{Raga2020b} time-dependence forms
a working surface (the ``head'' of the plasmon) followed
by the material in the low velocity, final wing of the
ejection pulse (forming the Hubble law ``tail''). These
authors called this flow the ``head/tail plasmon'', adapting
the name proposed by \citet{DeYoung1967} for
a clump-like outflow.

In the present paper, we study an alternative type of
``single pulse outflow'' that also produces a structure
with a Hubble law of linearly increasing velocities with
increasing distances from the outflow source, We propose
a cylindrical ejection with:
\begin{itemize}
\item a ``square pulse'' time-dependent ejection velocity, with
a sudden ``turning on'' at an ejection time $\tau=0$
and a ``turning off'' at $\tau=\tau_0$,
\item a parabolic initial cross section for the ejection
  velocity (with a peak, on-axis velocity and 0 velocity
  at the outer edge $r_j$).
\end{itemize}
This is in contrast to the single pulse outflows
studied by \citet{Raga2020a,Raga2020b}, who proposed parabolic
or Gaussian time-dependencies for the velocity
and a top-hat cross section ejection for the ejection.

The paper is organized as follows. In section 2, we describe
an analytic model, based on the ``center of mass'' formalism
of \citet{Canto2000}, which leads to a simple solution
for the motion of the working surface produced by the
(non-top hat cross section) ejection pulse. In section 3, we present
an axisymmetric numerical simulation (with parameters
appropriate for a high velocity not in a PN), and compare
the obtained results with the analytic models. Predictions
of position-velocity (PV) diagrams are done from the numerical
model. Finally, the results are discussed in section 4.

\section{The analytic model}

\subsection{The shape of the working surface}

Let us consider a hypersonic, cylindrical ejection with a
time-dependent, ``square pulse'' ejection velocity, and a non-top
hat cross section. The ejected material
will be free-streaming (because the pressure force is
negligible) until it reaches a leading working surface
(or ``head'') formed in the interaction between the outflow
and the surrounding environment. This situation is shown
schematically in Figure 1.

If the material in the working surface is locally well mixed,
the center of mass formalism of \citet{Canto2000} will
give the correct position $x_{cm}$ of the working surface
for all radii $r$ in the cross section of the outflow. Then,
$$  x_{cm}(r,t)= $$
\begin{equation}
  \frac{\int_0^\tau\rho_0(r,\tau')u_0(r,\tau')x_j(r,t,\tau')d\tau'
    +\int_0^{x_{cm}}\rho_a(x)xdx}{\int_0^\tau\rho_0(r,\tau')
    u_0(r,\tau')d\tau'+\int_0^{x_{cm}}\rho_a(x)dx}\,,
  \label{xcm}
\end{equation}
where $r$ is the cylindrical radius, $x$ is the distance
along the outflow axis,
$\rho_a(x)$ is the (possibly position-dependent)
ambient density, $\tau'$ is the time at which the flow parcels
were ejected. $u_0(r,\tau')$ and $\rho_0(r,\tau')$ are the
time-dependent velocity and density ejection cross sections
(respectively),
\begin{equation}
  x_j(r,t,\tau')=(t-\tau')u_0(r,\tau')\,,
  \label{xttau}
\end{equation}
is the position that the fluid parcels would have if
they were still in the free-flow regime and
$\tau$ is the time at which the parcels now (i.e., at
time $t$) entering the working surface were ejected.
This time $\tau$ can be found by appropriately
inverting the free-streaming flow relation:
\begin{equation}
  \frac{x_{cm}(r,t)}{t-\tau}=u_0(r,\tau)\,.
  \label{fflow}
\end{equation}

\begin{figure}[!t]
\includegraphics[width=\columnwidth]{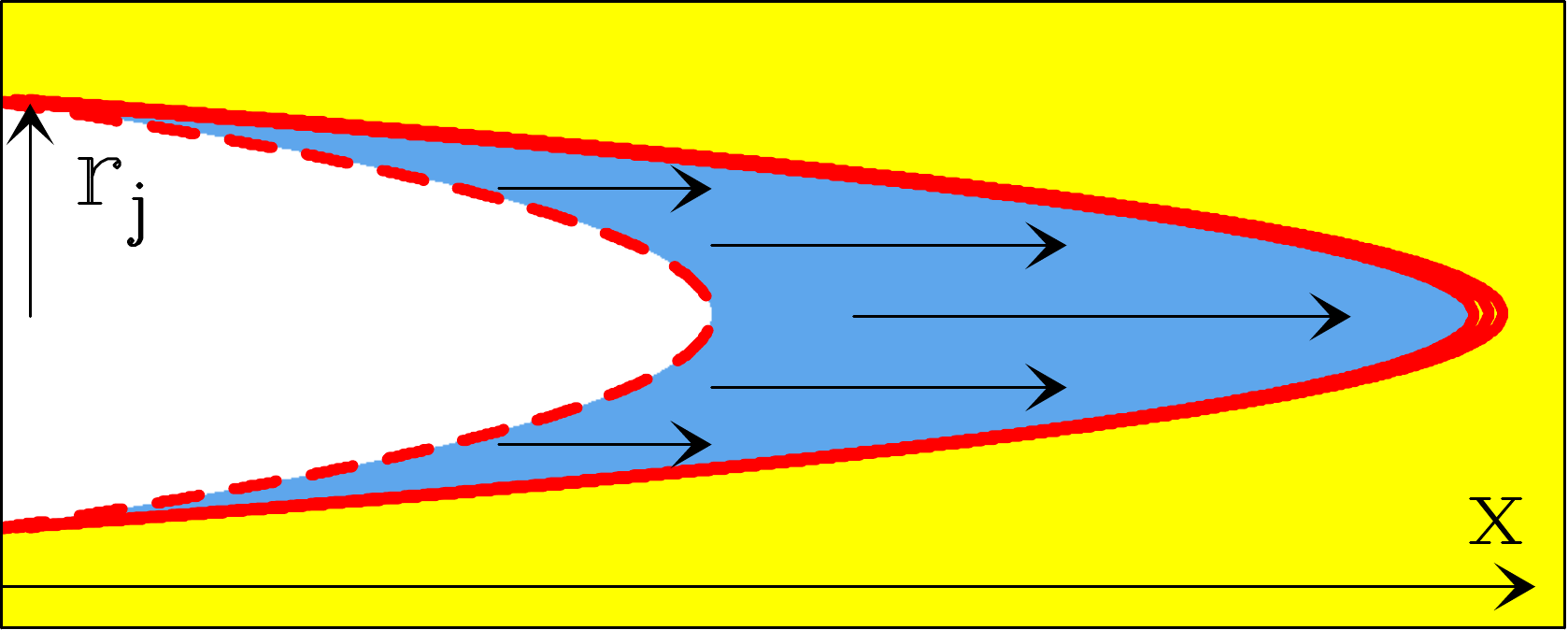}
\caption{Schematic diagram showing the interaction of
  an ejection pulse (of duration $\tau_0$ and initial
  radius $r_j$, travelling along the $x$-direction)
  with a non-top hat ejection velocity
  cross section interacting with a uniform environment.
  The outflow interacts with the environment forming a two-shock,
  curved working surface (thick, solid red curve). At evolutionary
  times $t>\tau_0$ the source is no longer ejecting material,
  and therefore an empty region (limited by the dashed,
  red curve) is formed close to the outflow source. At
  large enough times, all of the ejected material will join
  the working surface, and the empty region will be bounded
  by the bow shock.}
\label{fig1}
\end{figure}

Now, let us assume that we have an ejection pulse
with a velocity
\begin{equation}
  u_0(r,\tau)=v_0f(r)\,;\,\,\,\,0\leq\tau\leq\tau_0,
  \label{u0t}
\end{equation}
with constant $v_0$. For $\tau<0$ and $\tau>\tau_0$ there
is no ejection. The function $f(r)$ is the radial profile
of the ejection velocity, which we will assume has a peak
at $r=0$ and low velocities at the outer radius $r_j$
of the cylindrical ejection. We will furthermore assume
that the ejection density $\rho_0$ is time independent, and
that the outflow moves into a uniform environment of
density $\rho_a$

We now
introduce the ejection velocity given by equation (\ref{u0t})
and constant $\rho_0$ and $\rho_a$ (see above) in equations
(\ref{xcm}-\ref{xttau}) to obtain:
\begin{equation}
  \frac{\sigma}{2}x_{cm}^2+v_0f(r)\tau x_{cm}+
    v_0^2f^2(r)\tau\left(\frac{\tau}{2}-t\right)=0\,,
    \label{xcma}
\end{equation}
where
\begin{equation}
  \sigma\equiv \frac{\rho_a}{\rho_0}
\end{equation}
is the environment-to-outflow density ratio.
This equation can be inverted to obtain $x_{cm}$ as a function
of $t$ and $\tau$:
\begin{equation}
  x_{cm}=\frac{v_0f(r)\tau}{\sigma}\left[\sqrt{1+\frac{2\sigma}{\tau}
      \left(t-\frac{\tau}{2}\right)}-1\right]\,,
  \label{xcmb}
\end{equation}
where $\tau$ is the ejection time of the material entering
the working surface at an evolutionary time $t$ (see equation
\ref{fflow}).

Now, as $t$ grows, the ejection time $\tau$ also grows. and
eventually reaches $\tau_0$. For $\tau>\tau_0$, all of the
ejected material (at a given radius $r$) has fully entered
the working surface, and for larger times the position of the
working surface evolves following equation (\ref{xcmb}) with
$\tau=\tau_0$.

It is also possible to obtain $x_{cm}$ fully as a function of
evolutionary time $t$ by combining equations (\ref{fflow}) and
(\ref{u0t}) to obtain
\begin{equation}
  \tau=t-\frac{x_{cm}}{v_0f(r)},
    \label{tauxv}
\end{equation}
valid for $\tau\leq \tau_0$, and substituting this into
equation (\ref{xcma}). After some manipulation, one obtains:
\begin{equation}
  x_{cm}=\frac{v_0f(r)t}{\sigma^{1/2}+1}\,,
  \label{xcm1}
\end{equation}
which (not surprisingly) corresponds to the constant velocity
motion predicted from a simple ``ramp-pressure balance'' argument.
This solution was derived for the head of a constant velocity,
non-top hat cross section jet by \citet{Raga1998}.

For $\tau>\tau_0$, the position is given by equation (\ref{xcmb})
with $\tau=\tau_0$:
\begin{equation}
  x_{cm}=\frac{v_0f(r)\tau_0}{\sigma}\left[\sqrt{1+\frac{2\sigma}{\tau_0}
      \left(t-\frac{\tau_0}{2}\right)}-1\right]\,.
  \label{xcm2}
\end{equation}

The transition between the regimes of equation (\ref{xcm1}) and
(\ref{xcm2}) occurs at the evolutionary time $t_c$
when the material ejected at $\tau_0$
catches up with the working surface. The position of the last
ejected material is:
\begin{equation}
  x_0=(t-\tau_0)v_0f(r)\,,
  \label{tc0}
\end{equation}
and it catches up with the working surface when $t=t_c$ and $x_0=x_{ws}$.
We can now use the value of $x_{ws}$ obtained from equations (\ref{xcm1})
or (\ref{xcm2}), which when substituted in equation (\ref{tc0})
both lead to:
\begin{equation}
  t_c=\left(1+\sigma^{-1/2}\right)\tau_0\,,
  \label{tc}
\end{equation}
which is independent of $r$. Therefore, at a time $t_c$, the material
of the pulse ejected at all radii is fully incorporated into the
working surface. At a time $t_c$, the working surface has
a shape:
\begin{equation}
  x_{c}(r)=\frac{v_0f(r)\tau_0}{\sigma^{1/2}}\,,
  \label{xc}
\end{equation}
obtained combining equations (\ref{xcm1}) and (\ref{tc}).

\subsection{The velocity structure}

The velocity of the material within a fully mixed working surface
is directed along the $x$-axis (see Figure 1). The position-dependent
velocity can be straightforwardly obtained by calculating the
time-derivative of the $x_{cm}(r,t)$ locus of the working
surface (given by equations \ref{xcm1} and \ref{xcm2}, depending on
the value of $t$).

For $t\leq t_c$ (see equation \ref{tc}), from equation (\ref{xcm1})
we obtain:
\begin{equation}
  v_{cm}=\frac{v_0f(r)}{1+\sigma^{1/2}}=\frac{x_{cm}}{t}\,.
  \label{vcm1}
\end{equation}
Therefore, the velocity in the curved working surface has
a ``Hubble law'' of linearly increasing velocities as
a function of distance along the $x$-axis, with a slope
of $1/t$.

For $t>t_c$ (see equation \ref{tc}), from equation (\ref{xcm2})
we obtain:
$$
v_{cm}=\frac{v_0f(r)}{\sqrt{1+\frac{2\sigma}{\tau_0}
      \left(t-\frac{\tau_0}{2}\right)}}=
$$
\begin{equation}
\frac{\sigma x_{cm}}{\tau_0\sqrt{1+\frac{2\sigma}{\tau_0}
    \left(t-\frac{\tau_0}{2}\right)}
  \left[\sqrt{1+\frac{2\sigma}{\tau_0}
      \left(t-\frac{\tau_0}{2}\right)}-1\right]}\,.
\label{vcm2}
\end{equation}
Again, the velocity as a function of distance follows a linear,
``Hubble law''. The slope of this law (see equation \ref{vcm2})
is $1/t$ for $t=t_c$, and approaches a value of $1/(2t)$ for
$t\gg \tau_0$.
  
\subsection{Solutions for different $\sigma$ values}

If we choose values for the density ratio $\sigma=\rho_a/\rho_0$,
from equations (\ref{xcm1}-\ref{xcm2}) we obtain the position
$x_{ws}$ and from equations (\ref{vcm1}-\ref{vcm2}) the
velocity of the working surface on the symmetry axis. The
positions and velocities obtained for $\sigma=0$, 0.1, 0,5, 1.0 and 2.0
are shown in Figure 2.

For $\sigma=0$ (the ``free plasmon'') the plasmon head
moves at a constant velocity $v_0$ (see equation \ref{u0t}).
For $\sigma>0$, the working surface moves at a constant
velocity (given by equation \ref{vcm1}) for $t\leq t_c$ (see
equation \ref{tc}), and has a monotonically decreasing
velocity for $t>t_c$. The velocity at all times has lower
values for larger $\sigma$.

\begin{figure}[!t]
\includegraphics[width=\columnwidth]{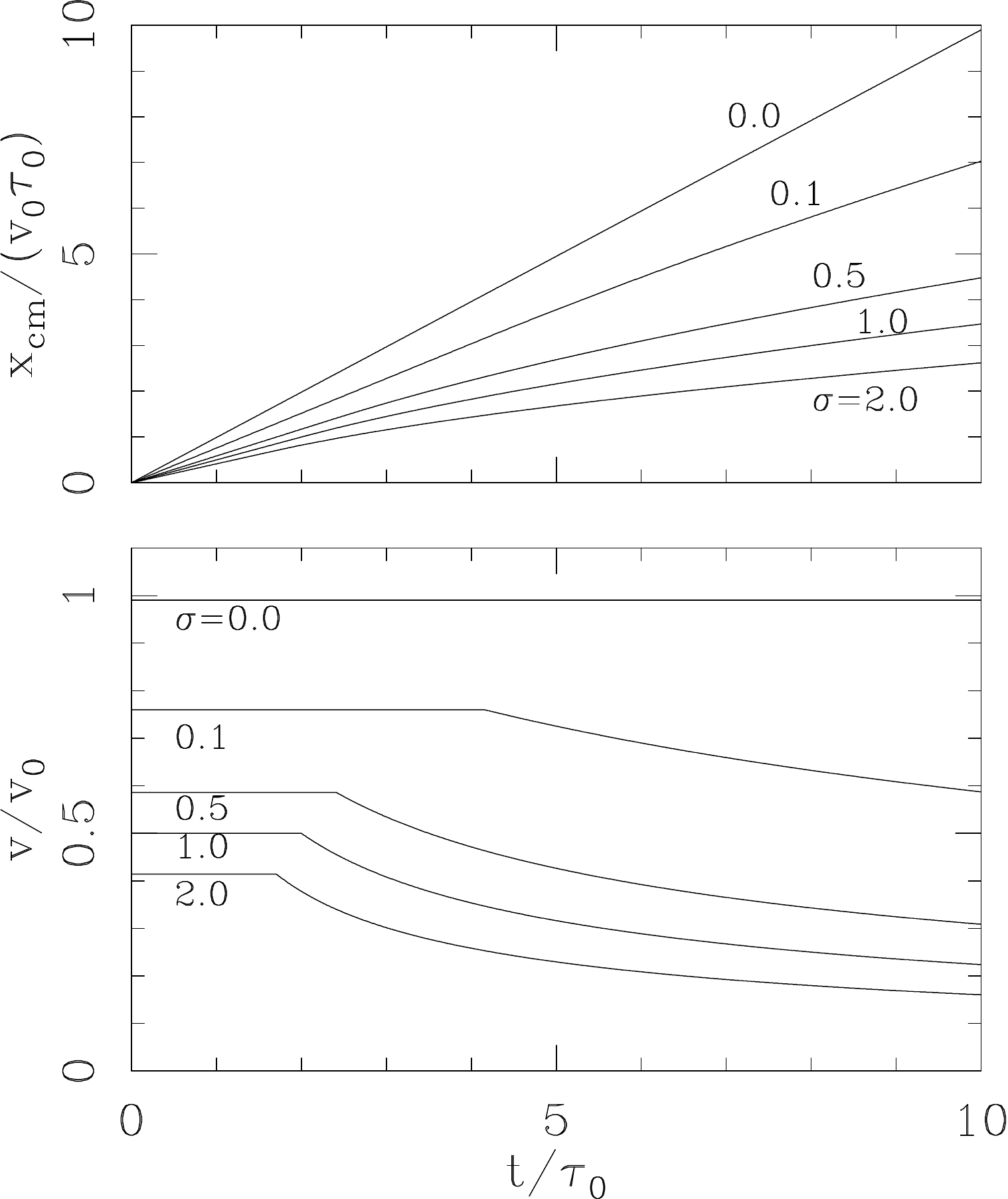}
\caption{Position (top) and axial velocity (bottom) of the
  head of the plasmon as a function of time. The curves
  are labelled with the values of $\sigma$ used to calculate
  the solutions (see equations \ref{xcm1}, \ref{xcm2}, \ref{vcm1}
  and \ref{vcm2}).}
\label{fig2}
\end{figure}

In order to illustrate the shapes that the plasmon (i.e., the
working surface) can take, we choose a parabolic ejection velocity
cross section (see equation \ref{u0t}):
\begin{equation}
  f(r)=1-\left(\frac{r}{r_j}\right)^2\,,
  \label{fr}
\end{equation}
where $r_j$ is the radius of the cylindrical outflow.

In Figure 3, we show the time-evolution of the flow for
three different values of the environment-to-ejection density
ratio: $\sigma=0$, 0.1 and 0.5. For $\sigma=0$, the time at which
the ejected material fully enters the working surface is
$t_c\to \infty$ (see equation \ref{tc}). For $\sigma=0.1$ and 0.5,
we obtain $t_c=4.1\tau_0$ and $2.41\tau_0$, respectively.
The shapes shown in Figure 3 were obtained using equation (\ref{xcm1})
for times $t\leq t_c$ and equation (\ref{xcm2}) for $t>t_c$ (this
case applies only to the $t=3$ and $4\tau_0$ frames of the
$\sigma=0.5$ case).

\begin{figure}[!t]
\includegraphics[width=\columnwidth]{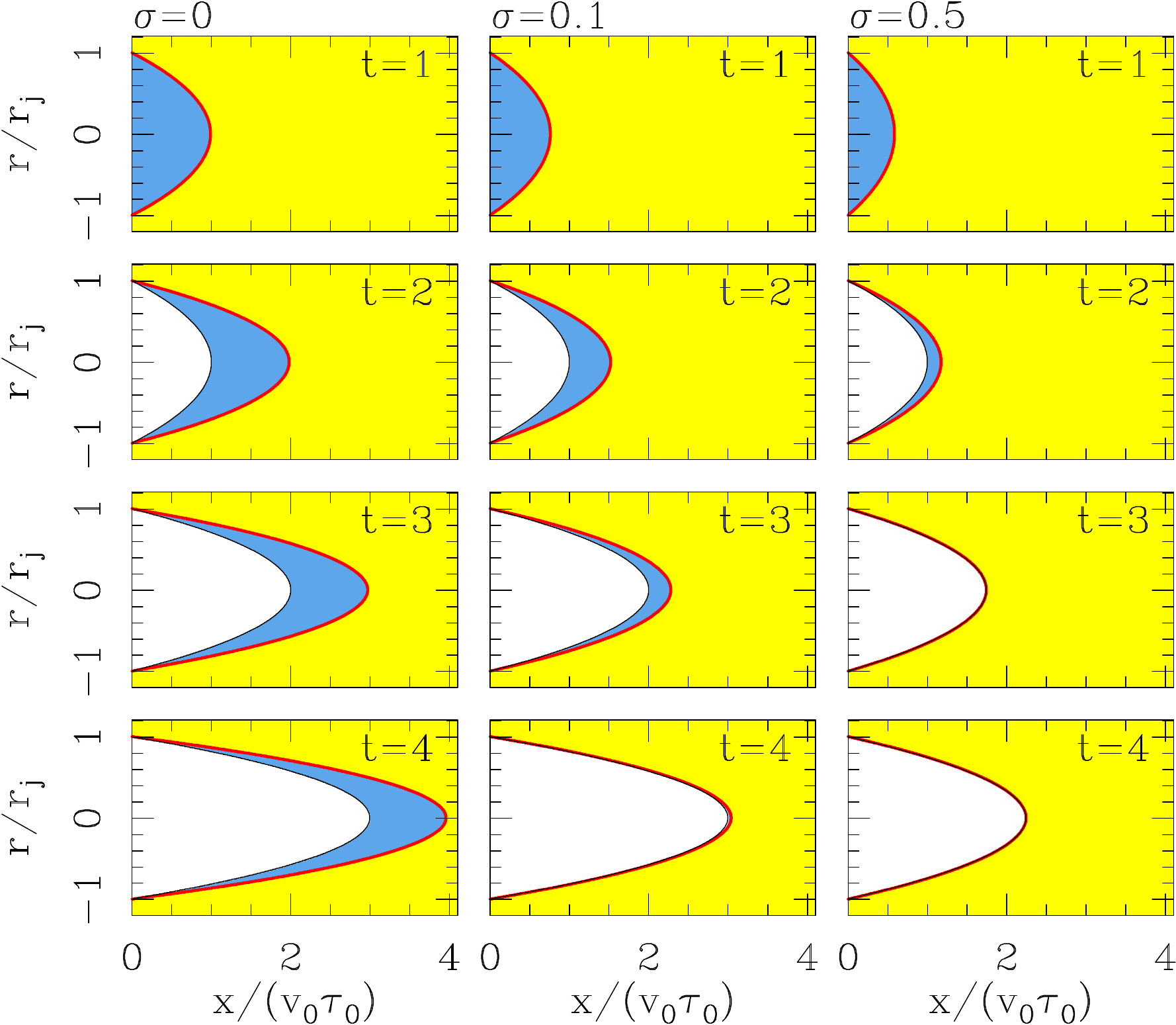}
\caption{Solutions for an outflow pulse with a parabolic
  ejection velocity cross section (see section 2.3). The three
  columns show the time-evolutions obtained for different values
  of $\sigma=\rho_a/\rho_0$, and are labelled with the corresponding
  $\sigma$ (at the top of the top graphs). The four lines correspond
  to different evolutionary times: $t=\tau_0$ (top), $2\tau_0$,
  $3\tau_0$ and $4\tau_0$ (bottom). The working surface is shown with the
  thick, solid line, and the ``empty cavity'' region is shaded white.
  The blue region (not always present) is the ejected material which
  has still not been incorporated into the working surface.}
\label{fig3}
\end{figure}

Also shown in Figure 3 is the ``empty region'' formed for
$t>\tau_0$ (i.e., when the ejection has already stopped) close
to the outflow source (see equation \ref{tc0}).
In the $\sigma=0$ case, the working
surface moves freely, and therefore the ejected material (shown
in blue in Figure 3) never catches up with it. In the $\sigma=0.1$
case, in the $t=4\tau_0$ frame most of the ejected material has
already caught up with the working surface, and in the
$\sigma=0.5$ case in the $t=3$ and $4\tau_0$ frames (which
have $t>t_c$, see above) all of the outflow material is within
the working surface, and the ``empty region'' fills the volume
between the outflow source and the working surface.

\section{A numerical simulation}

\subsection{Flow parameters}

In order to illustrate in more detail the full characteristics of the
flow, we compute an axisymmetric numerical simulation of the ``parabolic
cross section plasmon'' described in section 2.3 using the
{\sc walicxe}-2D code \citep{Esquivel2009}. We choose parameters
appropriate for a high velocity clump in a PN: an axial velocity
with an on-axis value $v_0=200$~km~s$^{-1}$ (decreasing parabolically
to zero at a radius $r_j$, see equation \ref{fr}),
an initial radius $r_j=10^{16}$~cm,
an ejection atom+ion number density $n_0=10^4$~cm$^{-3}$ (independent
of radius) and an
ambient density $n_a=100$~cm$^{-3}$. Initially, both the outflow
and the environment have a $10^4$~K temperature. The ejection is
imposed at $t=0$ (at the beginning of the simulation) and
ends at a time $\tau_0=100$~yr.
For these parameters, the environment to outflow
density ratio has a value $\sigma=0.1$, and
we then expect the ejected material to be
fully incorporated into the working surface at a time
$t_c=416.2$~yr (see equation \ref{tc}).

We assume that all of the flow is photoionized by the central star
of the PN. We consider this photoionization in an approximate way by
imposing a minimum temperature $T=10^4$~K and full ionization for
Hydrogen throughout the flow. The parametrized cooling function
of Biro \& Raga (1994) is used for $T>10^4$~K.

The computational domain has a size of $(35,\,8.75)\times 10^{16}$~cm
(along and across the outflow axis, respectively),
resolved with a 7-level binary adaptive grid with a maximum
resolution of $8.54\times 10 ^{13}$~cm. An inflow boundary is applied at
$x=0$ and $r>r_j$ for $t<\tau_0$, a reflection boundary is
applied outside the injection region (at $x=0$) and on the symmetry axis,
and a free outflow is imposed in the remaining grid boundaries.

\subsection{Results}

We have run the simulation described in section 3.1 for a total
time of 600~yr. Figures 4 and 5 show time-frames (at times
$t=100$, 200, 300, 400, 500 and 600~yr) of the resulting
density stratification. In these figures, we show
the shape of the working surface (equations \ref{xcm1} and \ref{xcm2}).
For times $t\leq t_c=416.2$~yr (see section 3.1), we also show
the inner edge of the ``empty cavity'' of the analytic model
(equation \ref{tc0}). For $t>t_c$, all of the region
inside the working surface is in the ``empty cavity'' regime,
and for $t\leq \tau_0=100$~yr there is no empty region.

\begin{figure}[!t]
\includegraphics[width=0.95\columnwidth]{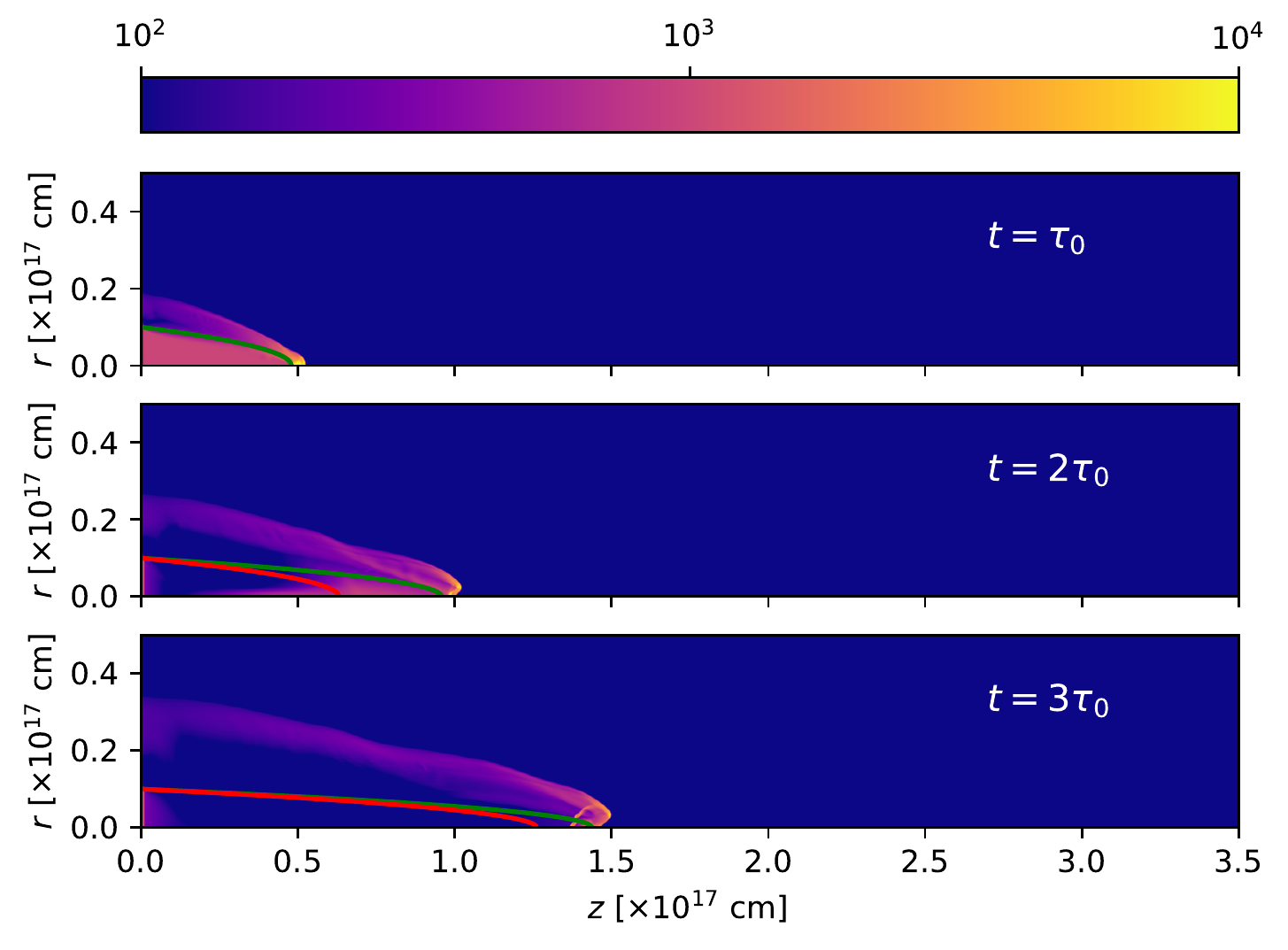}
\caption{Number density stratifications obtained from the
  numerical simulation for times $t=100$, 200 and 300~yr. The
  densities are shown with the logarithmic colour scale given by
  the top bar (in cm$^{-3}$). The shape of the working surface obtained
  from the analytic model is shown with the green curve, and the inner
  limit of the analytic ``empty cavity'' is shown with the red
  curve. The distances along and across the outflow axis are given
in units of $10^{17}$~cm.}
\label{fig4}
\end{figure}

\begin{figure}[!t]
\includegraphics[width=0.95\columnwidth]{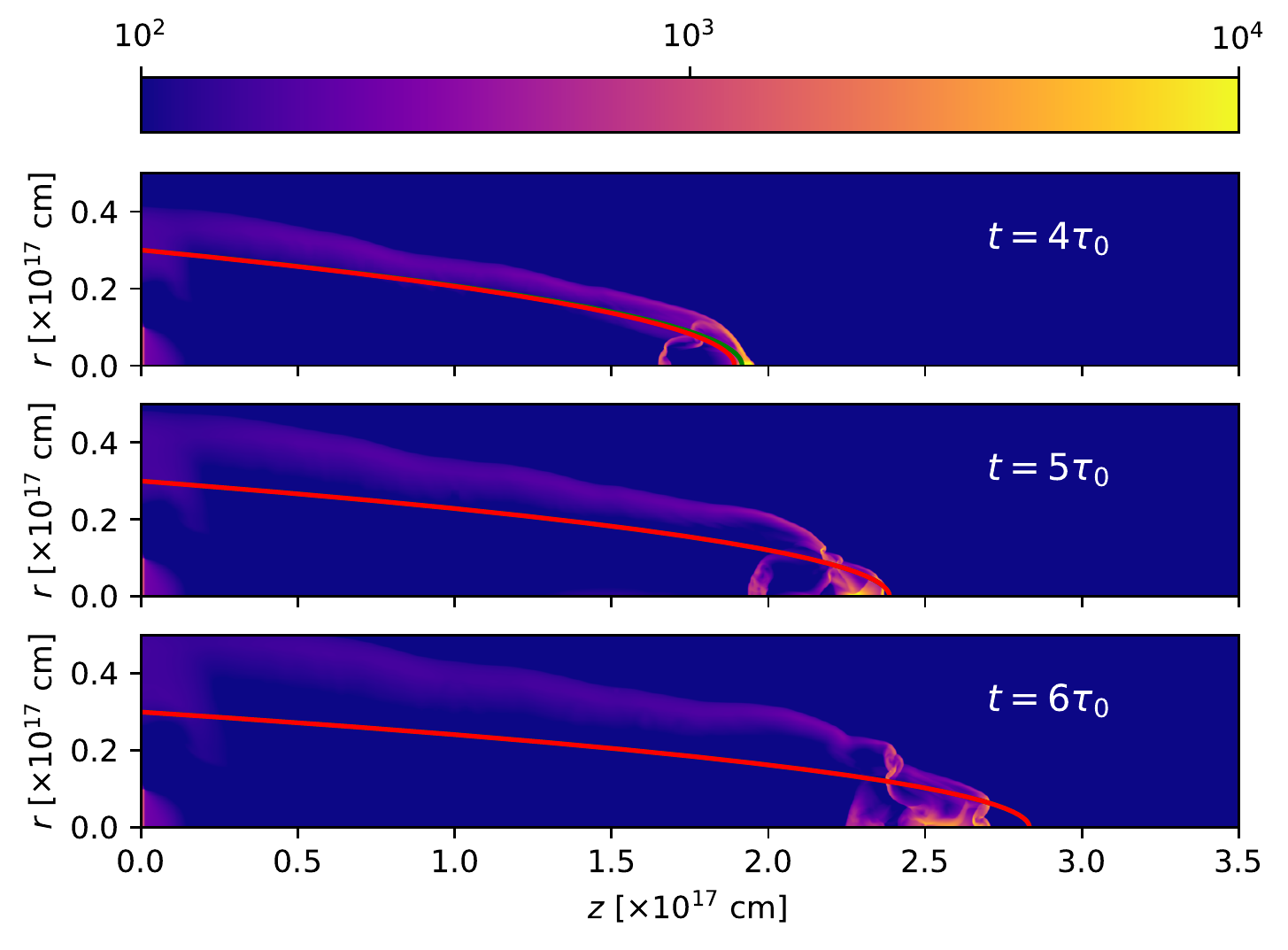}
\caption{The same as Figure 4, but for times $t=400$, 500
and 600~yr.}
\label{fig5}
\end{figure}

It is clear that even though at early times (see the $t=\tau_0$
frame of Figure 4) the working surface of the numerical simulation
has a shape that partially agrees with the analytic model, at later
times the working surface has bow shock wings which are considerably
broader than the analytic prediction (see the remaining frames
of Figures 4 and 5). This difference is partly due to the lack of
perfect mixing (assumed in the analytic model) between
outflow and environment material in the numerical simulation.
The other effect that pushes out material sideways from the
head of the working surface is the radial gas pressure
gradient (also not included in the analytic model). However,
the position of leading region of the working surface approximately
agrees with the analytic model at all times (see Figures 4 and 5).

We have computed the recombination cascade H$\alpha$ emission coefficient,
and integrated it through lines of sight in order to compute
intensity maps. Figures 6 and 7 show the emission
maps computed assuming a $30^\circ$ angle between the outflow
axis and the plane of the sky, for times
$t=100$, 200, 300, 400, 500 and 600~yr.

From Figures 6 and 7, we see that
the earlier maps (the $t=100$ and 200~yr, top two
frames of Figure 6) show the emission from the ejected
material before it reaches the working surface). In all of
the later maps, we see a bright, compact component in the
leading, on-axis region of the working surface, and the
emission of extended bow shock wings trailing this
clump.

\begin{figure}[!t]
\includegraphics[width=0.95\columnwidth]{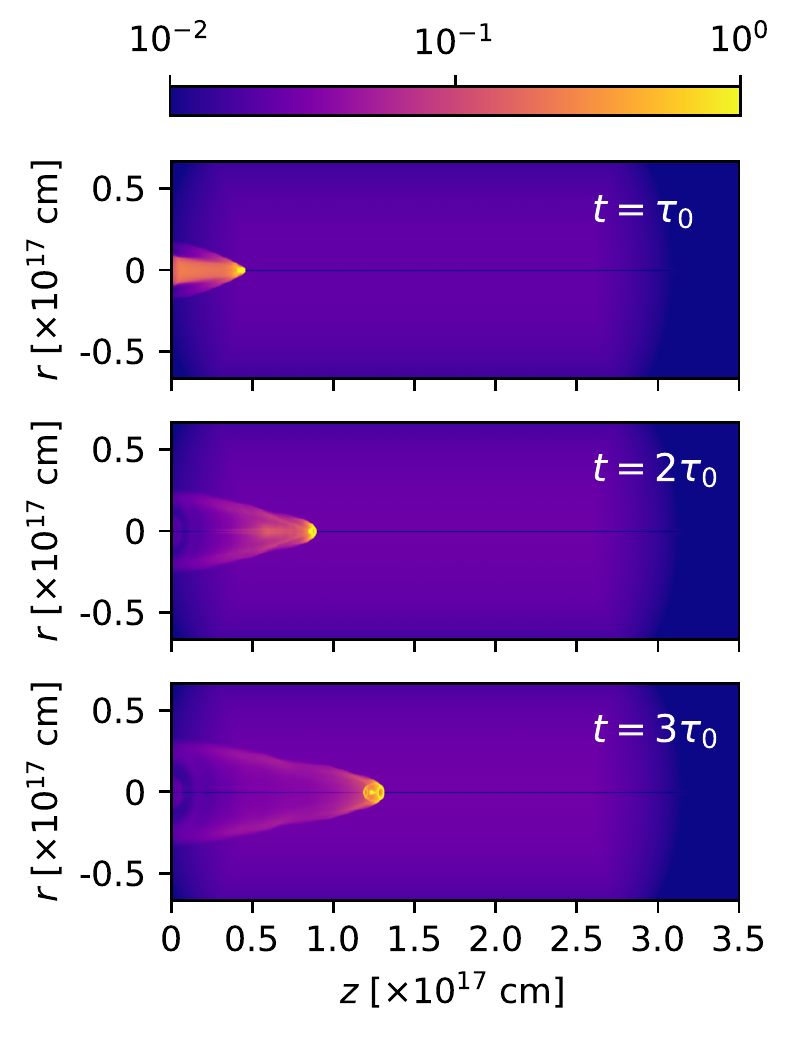}
\caption{H$\alpha$ maps obtained from the
  numerical simulation for times $t=100$, 200 and 300~yr.
  The maps are computed assuming a $30^\circ$ angle between
  the outflow axis and the plane of the sky. The
  emission (normalized to the peak emission of each map)
  is shown with the logarithmic colour scale given by
  the top bar. The distances along and across the outflow axis are given
  in units of $10^{17}$~cm.}
\label{fig6}
\end{figure}

\begin{figure}[!t]
\includegraphics[width=0.95\columnwidth]{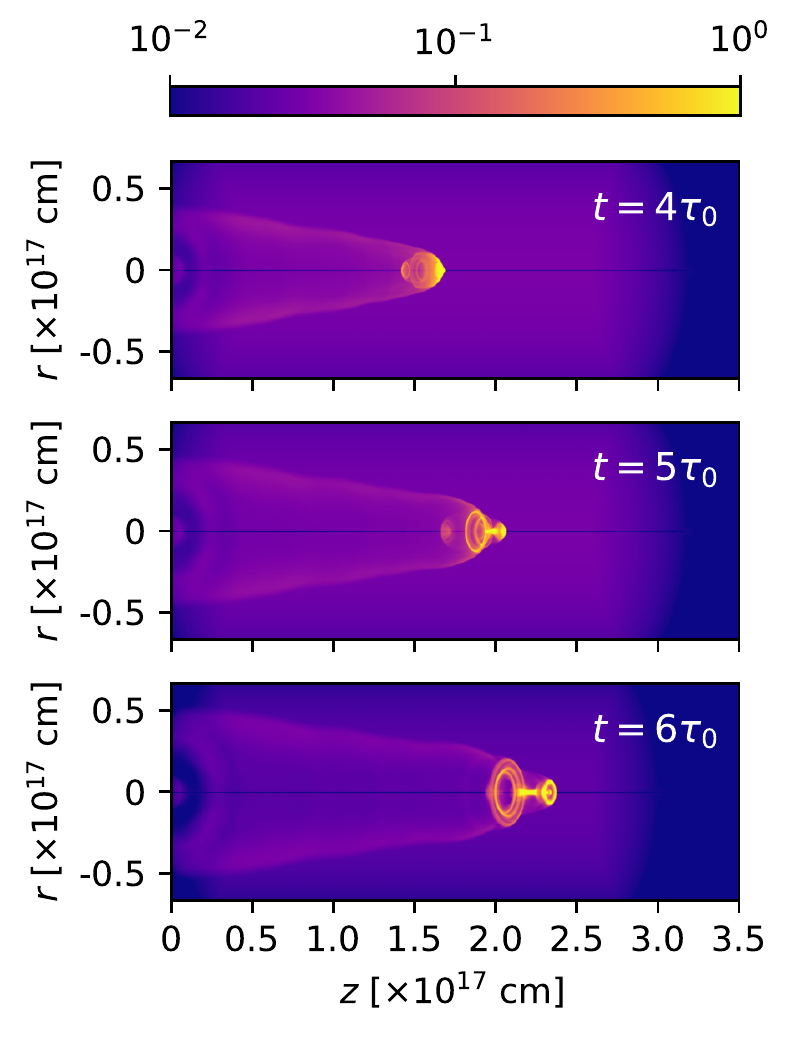}
\caption{The same as Figure 6, but for times $t=400$, 500
and 600~yr.}
\label{fig7}
\end{figure}

With the H$\alpha$ emission coefficient we have also
computed predicted position-velocity (PV) diagrams. These
PV diagrams correspond to long-slit spectra obtained with
a ``narrow'' spectrograph slit with a full projected width
of $2\times 10^{16}$~cm straddling the outflow axis (see Figures
8 and 9) and with a ``wide'' spectrograph
slit that incudes
all of the emission of the bow shock (see Figures 10 and 11), and
show the emission as a function of position along the outflow
axis and radial velocity (along the line of sight). Figures
(8,9) and (10.11)
show the PV diagrams computed for a $30^\circ$ orientation
of the outflow axis with respect to the plane of the sky, and for times
$t=100$, 200, 300, 400, 500 and 600~yr.

\begin{figure}[!t]
\includegraphics[width=0.95\columnwidth]{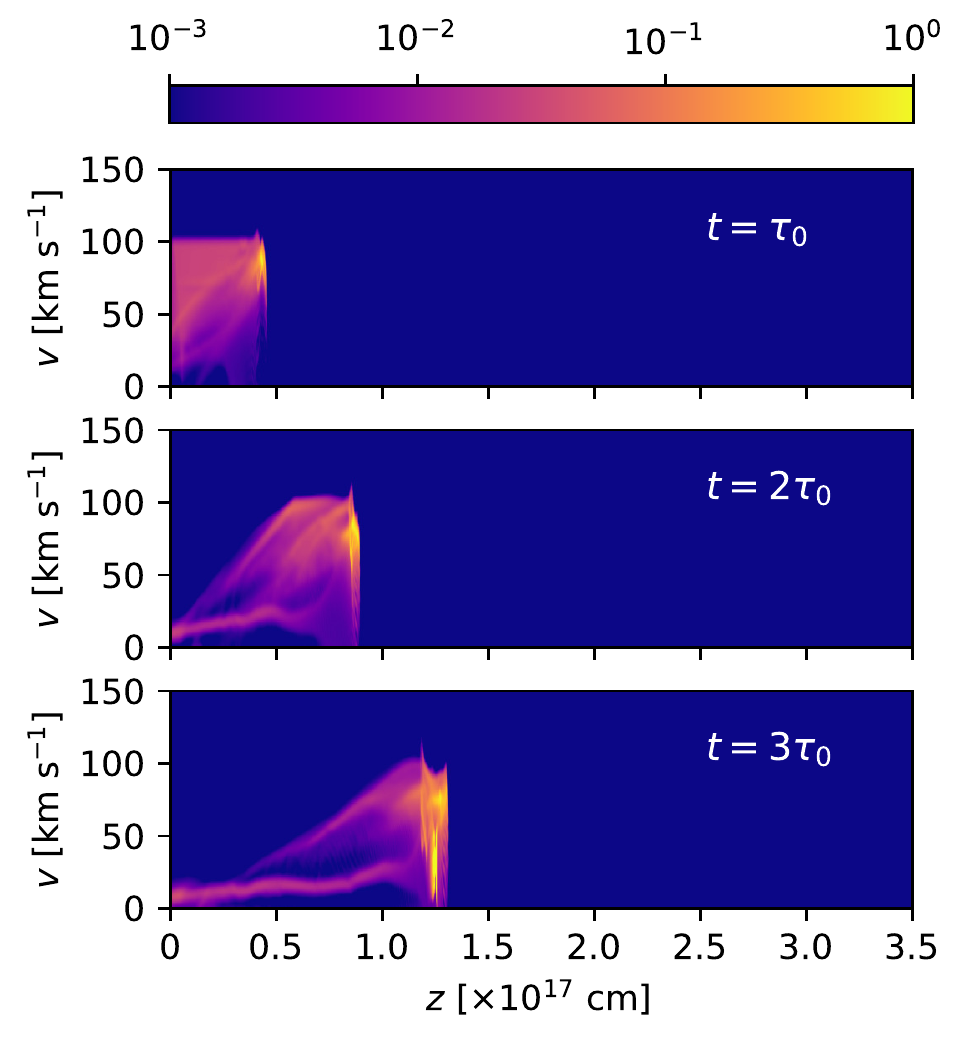}
\caption{H$\alpha$ position-velocity diagrams obtained from the
  numerical simulation for times $t=100$, 200 and 300~yr. These
  PV diagrams have been calculated assuming that a long spectrograph
  slit with a projected full width of $2\times 10^{16}$~cm straddles
  the symmetry axis of the flow.
  The maps are computed assuming a $30^\circ$ angle between
  the outflow axis and the plane of the sky. The
  emission (normalized to the peak emission of each map)
  is shown with the logarithmic colour scale given by
  the top bar. The distances along the outflow axis are given
  in units of $10^{17}$~cm, and the radial velocities in km~s$^{-1}$.}
\label{fig8}
\end{figure}

\begin{figure}[!t]
\includegraphics[width=0.95\columnwidth]{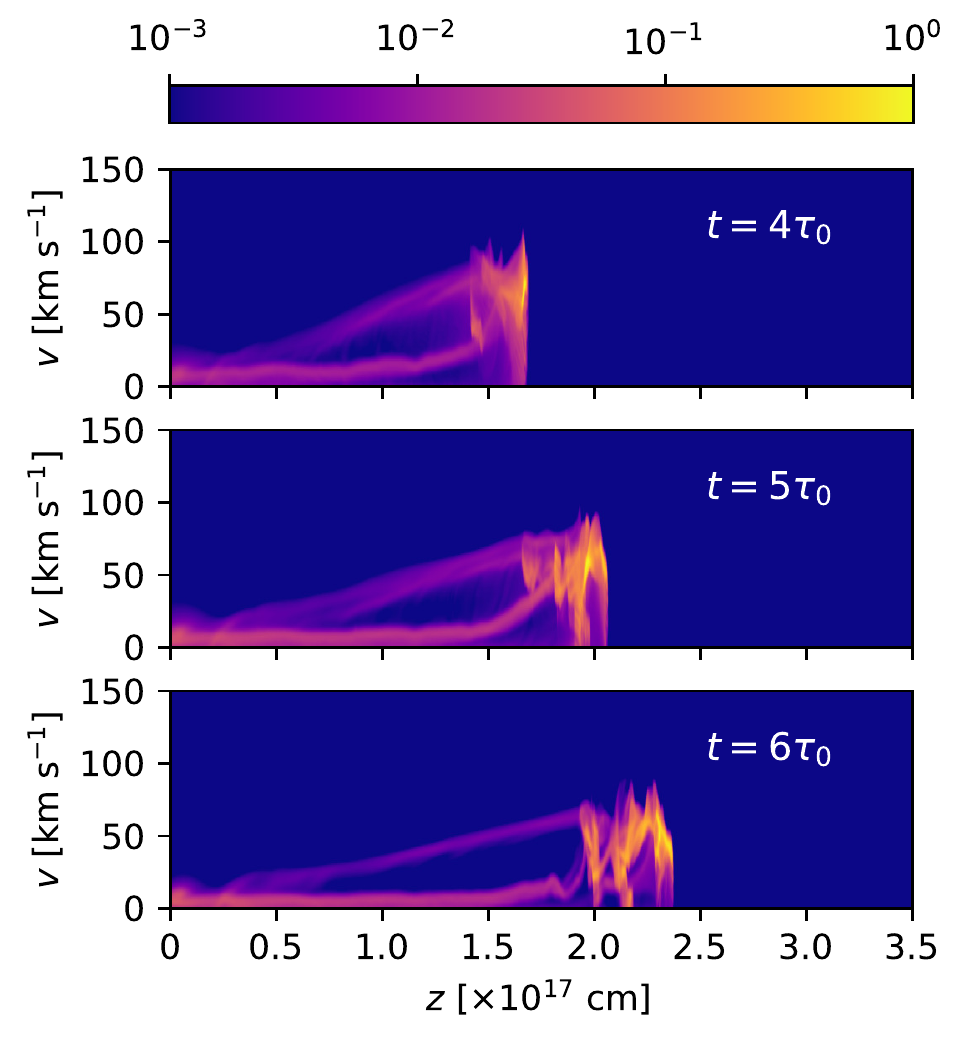}
\caption{The same as Figure 8, but for times $t=400$, 500
and 600~yr.}
\label{fig9}
\end{figure}

\begin{figure}[!t]
\includegraphics[width=0.95\columnwidth]{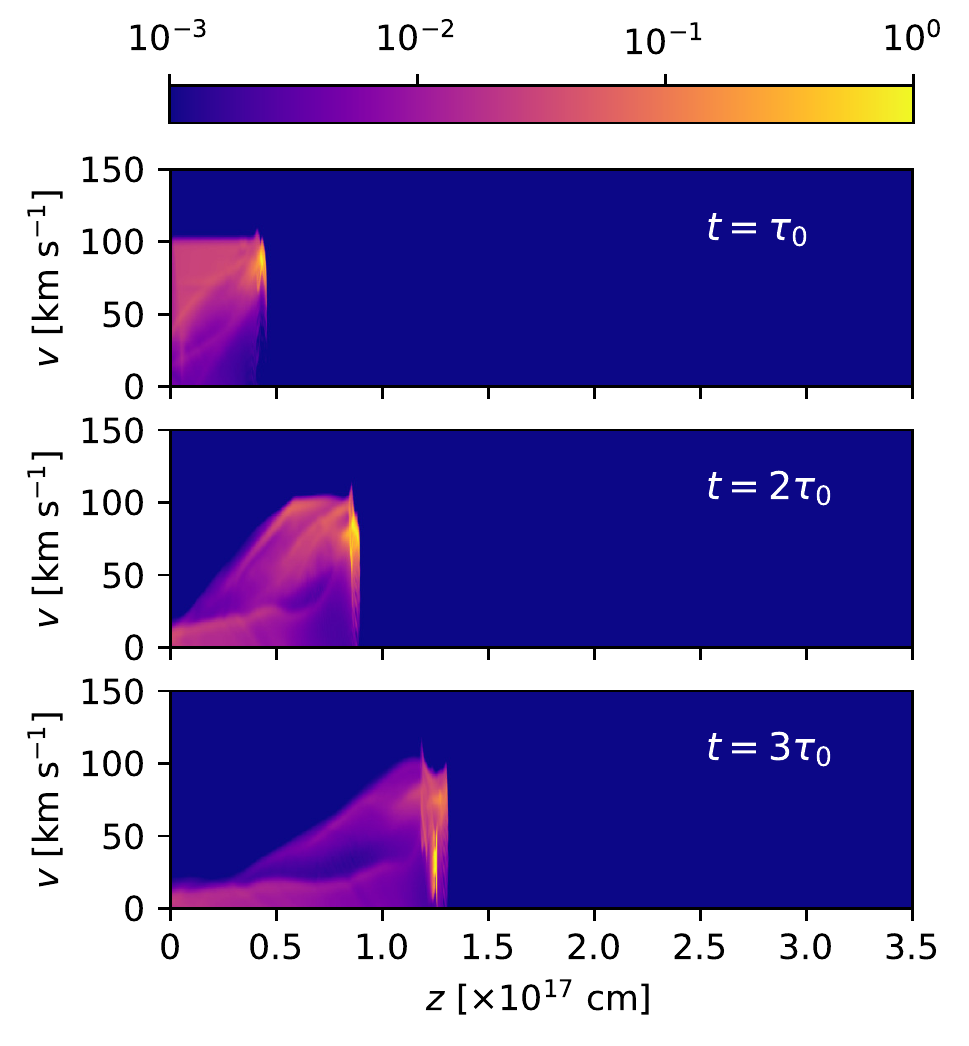}
\caption{The same as Figure 8, but with PV diagrams calculated
  for a wide spectrograph slit that straddles the symmetry
  axis and includes all of the emitting region of the flow.}
\label{fig10}
\end{figure}

\begin{figure}[!t]
\includegraphics[width=0.95\columnwidth]{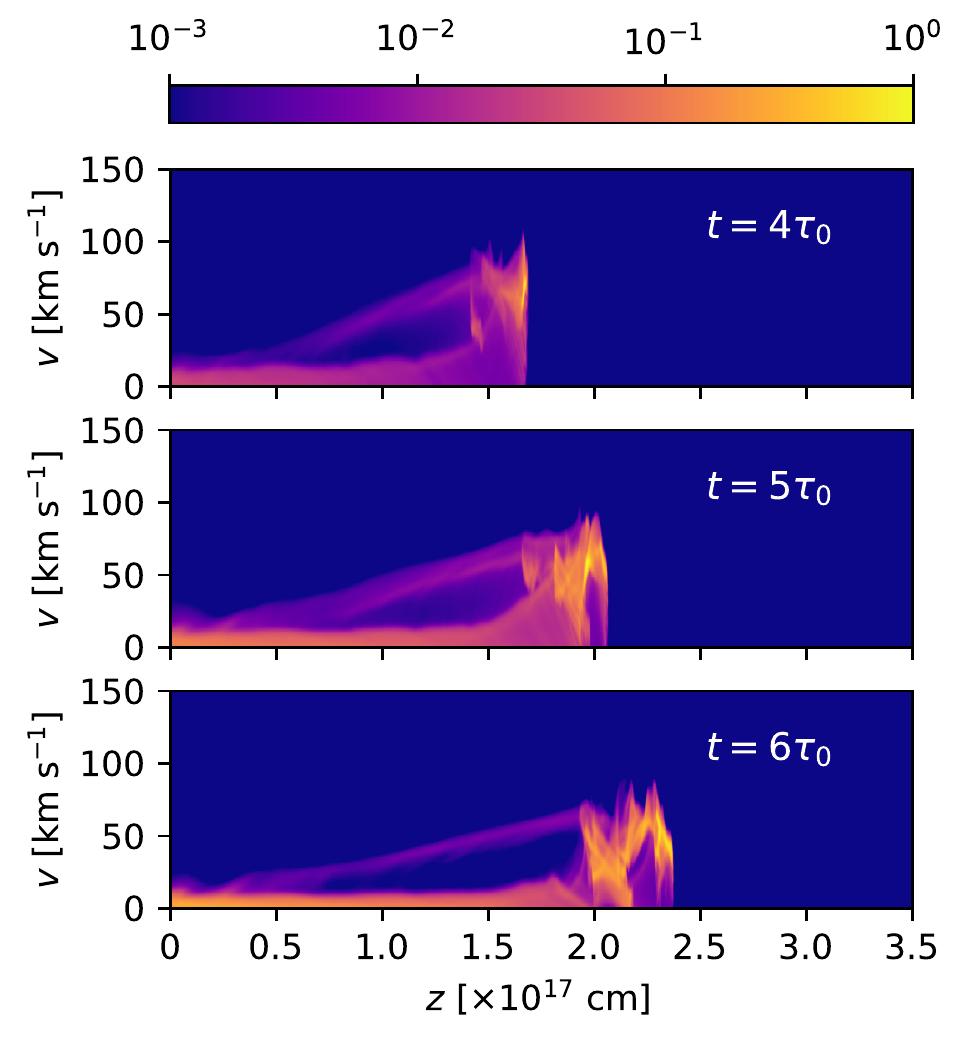}
\caption{The same as Figure 9, but with PV diagrams
  obtained with a wide spectrograph slit that includes all of
  the emitting region of the flow.}
\label{fig11}
\end{figure}

From these figures it is clear that in all of the PV diagrams
we see:
\begin{itemize}
\item qualitatively very similar results for slits of different
  widths (seen comparing Figure 8 to Figure 10, and 9 to 11),
  \item a bright, compact emission feature at the position and velocity
    of the on-axis, leading region of the working surface,
  \item an approximately linear ramp of increasing radial velocities,
    ending at the position of the leading clump.
\end{itemize}
Apart from these two components, in the earlier frames ($t=100$
and 200~yr, two top frames of Figure 8) we see the
ejected material (at a projected velocity of 100~km~s$^{-1}$)
before it reaches the working surface. This component disappears
at later times, since all of the ejected material has then been
incorporated into the working surface. Also, at all times
we see a low velocity component, which corresponds to
environmental material that has been shocked by the far
bowshock wings and has not mixed with the rest of the flow.
    
\section{Summary}

We have studied the flow resulting from a constant
density, collimated, cylindrical
non-top hat cross section ejection of material over a finite time $\tau_0$.
We first calculate an analytic model (based on the ``center
of mass formalism'' of \citealt{Canto2000}) with which we
obtain analytic expressions for the time-evolution of the
working surface produced by the interaction of the ejection
with a uniform environment.

This solution has two regimes:
\begin{itemize}
\item a working surface which is being fed by the
  ejected material (see equation \ref{xcm1}). This solution
  was previously derived by Raga et al. (1998),
\item a working surface in which all of the ejected material
  has already been incorporated (see equation \ref{xcm2}).
\end{itemize}
The transition between the two regimes occurs at the time
$t_c$ given by equation (\ref{tc}). For $t<t_c$, the region
inside the working surface is partly filled by the ejected
material (with an inner cavity with a boundary given
by equation \ref{tc0}). For $t>t_c$, the region
within the working surface is ``empty'' (i.e., in the
ballistic analytic model, see Figure 3).

For $t<t_c$, the working surface moves at a constant velocity,
and for $t>t_c$ it slows down, more strongly for larger
values of the environment-to-ejection density
ratio $\sigma=\rho_a/\rho_0$ (see Figure 2).
For these two regimes, we find that the material in the working
surface has a linear velocity vs. $x$ (the position along the
outflow axis) dependence, given by equations (\ref{vcm1}) and
(\ref{vcm2}).

We also compute an axisymmetric numerical simulation, with
conditions appropriate for a high velocity clump in a PN. We
find that the density structure initially shows a working
surface and a low density cavity that agree well with the
analytic predictions (see Figure 4). However, at later times
the numerical working surface develops bow shock wings that
are considerably broader than the ones of the analytic
prediction (see Figures 4 and 5). The position of the leading
region of the working surface shows a reasonably good agreement
with the analytic model for all of the computed times.

From the numerical simulation, we have calculated H$\alpha$
maps (Figures 6 and 7) and PV diagrams (Figures 8 to 11). We
find that the PV diagrams do show the linear radial velocity
vs. position ``Hubble law'' predicted from the analytic models
(see Figures 8 and 9).

Therefore, we have found a new way of straightforwardly obtaining
clump-like outflows with a ``Hubble law'' linear radial
velocity ramp joining them to the outflow source. This
is an alternative scenario to the one of the ``single
peak radial velocity pulse'' model of \citet{Raga2020a,Raga2020b},
which also produces ``Hubble law clumps''. Clearly, these two
possibilities are useful as guidelines to obtaining detailed
models of structures with these characteristics in
PN (see, e.g., \citealt{Dennis2008}) or in outflows in
star formation regions (see, e.g., \citealt{Zapata2020}).

We end by noting that the results presented in this paper
directly depend on quite arbitrary assumptions of a pulse-like
ejection and a non-top hat ejection velocity cross section. Reasonable
arguments for these two assumtions can be presented:
\begin{itemize}
\item an ejection with
  a limited duration is partially justified by the observation of clump-like
  flows in outflows from young and evolved stsars, which most likely
  imply such a time-limited ejection,
\item a non-top hat outlfow cross section could be the result of a
  magnetocentrifugal ejection from an accretion disk (which produces
  higher outflow velocities from the inner regions of the disk),
  or the result of an initial, turbulent outflow region that generates
  the centrally peaked velocity profile.
\end{itemize}
This is by no means concrete proof that the characetirstics
that we have assumed for the outflow are correct. This type of uncertainty
is present in the vast majority of the jet models in the astrophysical
literature, many of which share the assumption of a simple but unlikely
``sudden turn-on'', top hat cross section'' ejection.
    
\vfill\eject

\acknowledgments
This work was supported by the DGAPA (UNAM) grant IG100218 and IA103121.
A.C.R. acknowledges support from a DGAPA-UNAM postdoctoral fellowship. 
\vfill\eject

\vfill\eject

\end{document}